# Driven coherent oscillations of a single electron spin in a quantum dot

F. H. L. Koppens[1], C. Buizert[1], K. J. Tielrooij[1], I. T. Vink[1], K. C. Nowack[1], T. Meunier[1], L. P. Kouwenhoven[1] & L. M. K. Vandersypen[1]

[1]Kavli Institute of NanoScience, Delft University of Technology, PO Box 5046, 2600 GA, Delft, The Netherlands.

**The ability to control the quantum state of a single electron spin in a quantum dot is at the heart of recent developments towards a scalable spin-based quantum computer. In combination with the recently demonstrated exchange gate between two neighbouring spins, driven coherent single spin rotations would permit universal quantum operations. Here, we report the experimental realization of single electron spin rotations in a double quantum dot. First, we apply a continuous-wave oscillating magnetic field, generated on-chip, and observe electron spin resonance in spin-dependent transport measurements through the two dots. Next, we coherently control the quantum state of the electron spin by applying short bursts of the oscillating magnetic field and observe about eight oscillations of the spin state (so-called Rabi oscillations) during a microsecond burst. These results demonstrate the feasibility of operating single-electron spins in a quantum dot as quantum bits.**

The use of quantum mechanical superposition states and entanglement in a computer can theoretically solve important mathematical and physical problems much faster than classical computers[1,2]. However, the realization of such a quantum computer represents a formidable challenge, because it requires fast and precise control of fragile quantum states. The prospects for accurate quantum control in a scalable system are thus being explored in a rich variety of physical systems, ranging from nuclear magnetic resonance and ion traps to superconducting devices[3].

Electron spin states were identified early on as an attractive realization of a quantum bit[4], because they are relatively robust against decoherence (uncontrolled interactions with the environment). Advances in the field of semiconductor quantum dots have made this system very fruitful as a host for the electron spin. Since Loss and DiVincenzo's proposal[5] on electron spin qubits in quantum dots in 1998, many of the elements necessary for quantum computation have been realized experimentally. It is now routine to isolate with certainty a single electron in each of two coupled quantum dots[6–9]. The spin of this electron can be reliably initialized to the ground state, spin-up, via optical pumping[10] or by thermal equilibration at sufficiently low temperatures and strong static magnetic fields (for example, $T$=100 mK and $B_{ext}$=1 T). The spin states are also very long-lived, with relaxation times of the order of milliseconds[11–13].

Furthermore, a lower bound on the spin coherence time exceeding 1 μs was established, using spin-echo techniques on a two-electron system[14]. These long relaxation and coherence times are possible in part because the magnetic moment of a single electron spin is so weak. On the other hand, this property makes read-out and manipulation of single spins particularly challenging. By combining spin-to-charge conversion with real-time single-charge detection[15–17], it has nevertheless been possible to accomplish single-shot read-out of spin states in a quantum dot[13,18].

The next major achievement was the observation of the coherent exchange of two electron spins in a double dot system, controlled by fast electrical switching of the tunnel coupling between the two quantum dots[14]. Finally, free evolution of a single electron spin about a static magnetic field (Larmor precession) has been observed, via optical pump–probe experiments[19,20]. The only missing ingredient for universal quantum computation with spins in dots remained the demonstration of driven coherent spin rotations (Rabi oscillations) of a single electron spin.

The most commonly used technique for inducing spin flips is electron spin resonance (ESR)[21]. ESR is the physical process whereby electron spins are rotated by an oscillating magnetic field $B_{ac}$ (with frequency $f_{ac}$) that is resonant with the spin precession frequency in an external magnetic field $B_{ext}$, oriented perpendicularly to $B_{ac}$ ($hf_{ac}=g\mu_B B_{ext}$, with $\mu_B$ the Bohr magneton and g the electron spin g-factor). Magnetic resonance of a single electron spin in a solid has been reported in a few specific cases[22–24], but has never been realized in semiconductor quantum dots. Detecting ESR in a single quantum dot is conceptually simple[25], but experimentally difficult to realize, as it requires a strong, high-frequency magnetic field at low temperature, while accompanying alternating electric fields must be minimized. Alternative schemes for driven rotations of a spin in a dot have been proposed, based on optical excitation[26] or electrical control[27–29], but this is perhaps even more challenging and has not been accomplished either.

Here, we demonstrate the ability to control the spin state of a single electron confined in a double quantum dot via ESR. In a double dot system, spin-flips can be detected through the transition of an electron from one dot to the other[30,31] rather than between a dot and a reservoir, as would be the case for a single dot. This has the advantage that there is no need for the electron spin Zeeman splitting (used in a single dot for spin-selective tunnelling) to exceed the temperature of the electron reservoirs (~100 mK; the phonon temperature was ~40 mK). The experiment can thus be performed at a smaller static magnetic field, and consequently with lower, technically less demanding, excitation frequencies. Furthermore, by applying a large bias voltage across the double dot, the spin detection can be made much less sensitive to electric fields than is possible in the single-dot case (electric fields can cause photon-assisted tunnelling; see Supplementary Discussion). Finally, in a double dot, single-spin operations can in future experiments be combined with two-qubit operations to realize universal quantum gates[5], and with spin read-out to demonstrate entanglement[32,33].

**Device and ESR detection concept**

Two coupled semiconductor quantum dots are defined by surface gates (Fig. 1a) on top of a two-dimensional electron gas. By applying the appropriate negative voltages to the gates the dots can be tuned to the few-electron regime[8]. The oscillating magnetic field that drives the spin transitions is generated by applying a radio-frequency (RF) signal to an on-chip coplanar stripline (CPS) which is terminated in a narrow wire, positioned near the dots and separated from the surface gates by a 100-nm-thick dielectric (Fig. 1b). The current through the wire generates an oscillating magnetic field $B_{ac}$ at the dots, perpendicular to the static external field $B_{ext}$ and slightly stronger in the left dot than in the right dot (see Supplementary Fig. S1).

To detect the ESR-induced spin rotations, we use electrical transport measurements through the two dots in series in the spin blockade regime where current flow depends on the relative spin state of the electrons in the two dots[30,34]. In brief, the device is operated so that current is blocked owing to spin blockade, but this blockade is lifted if the ESR condition ($hf_{ac}=g\mu_B B_{ext}$) is satisfied.

This spin blockade regime is accessed by tuning the gate voltages such that one electron always resides in the right dot, and a second electron can tunnel from the left reservoir to the left dot (Fig. 1c and Supplementary Fig. S2). If this electron forms a double-dot singlet state with the electron in the right dot (S=↑↓−↓↑; normalization omitted for brevity), it is possible for the left electron to move to the right dot, and then to the right lead (leaving behind an electron in the right dot with spin ↑ or spin ↓), since the right dot singlet state is energetically accessible. If, however, the two electrons form a double-dot triplet state, the left electron cannot move to the right dot because the right dot's triplet state is much higher in energy. The electron also cannot move back to the lead and therefore further current flow is blocked as soon as any of the (double-dot) triplet states is formed.

## Role of the nuclear spin bath for ESR detection

In fact, the situation is more complex, because each of the two spins experiences a randomly oriented and fluctuating effective nuclear field of ~1–3 mT (refs 35, 36). This nuclear field, $B_N$, arises from the hyperfine interaction of the electron spins with the Ga and As nuclear spins in the host material, and is in general different in the two dots, by $\Delta B_N$. At zero external field and for sufficiently small double dot singlet–triplet splitting (see Supplementary Fig. S2d), the inhomogeneous component of the nuclear field causes all three triplet states ($T_0$, $T_+$ and $T_-$) to be admixed with the singlet S (for example, $T_0$=↑↓+↓↑ evolves into S=↑↓−↓↑ due to $\Delta B_{N,z}$, and $T_+$=↑↑ and $T_-$=↓↓ evolve into S owing to $\Delta B_{N,x}$). As a result, spin blockade is lifted. For $B_{ext} \gg \sqrt{\langle B_N^2 \rangle}$, however, the $T_+$ and $T_-$ states split off in energy, which makes hyperfine-induced admixing between $T_\pm$ and S ineffective ($T_0$ and S remain admixed; see Fig. 2a). Here spin blockade does occur, whenever a state with parallel spins (↑↑ or ↓↓) becomes occupied.

ESR is then detected as follows (see Fig. 1c). An oscillating magnetic field resonant with the Zeeman splitting can flip the spin in the left or the right dot. Starting from ↑↑ or ↓↓, the spin state then changes to ↑↓ (or ↓↑). If both spins are flipped, transitions occur between ↑↑ and ↓↓ via the intermediate state $\frac{\uparrow \pm \downarrow}{\sqrt{2}} \frac{\uparrow \pm \downarrow}{\sqrt{2}}$. In both cases, states with anti-parallel spins ($S_z$=0) are created owing to ESR. Expressed in the singlet-triplet measurement basis, ↑↓ or ↓↑ is a superposition of the $T_0$ and S state (↑↓=$T_0$+S). For the singlet component of this state, the left electron can transition immediately to the right dot and from there to the right lead. The $T_0$ component first evolves into a singlet due to the nuclear field and then the left electron can move to the right dot as well. Thus whenever the spins are anti-parallel, one electron charge moves through the dots. If such transitions from parallel to anti-parallel spins are induced repeatedly at a sufficiently high rate, a measurable current flows through the two dots.

**ESR spectroscopy**

The resonant ESR response is clearly observed in the transport measurements as a function of magnetic field (Fig. 2a, b), where satellite peaks develop at the resonant field $B_{ext}=\pm hf_{ac}/g\mu_B$ when the RF source is turned on (the zero-field peak arises from the inhomogeneous nuclear field, which admixes all the triplets with the singlet[36,37]). The key signature of ESR is the linear dependence of the satellite peak location on the RF frequency, which is clearly seen in the data of Fig. 2c, where the RF frequency is varied from 10 to 750 MHz. From a linear fit through the top of the peaks we obtain a *g*-factor with modulus 0.35±0.01, which lies within the range of reported values for confined electron spins in GaAs quantum dots[11,38–40]. We also verified explicitly that the resonance we observe is magnetic in origin and not caused by the electric field that the CPS generates as well; negligible response was observed when RF power is applied to the right side gate, generating mostly a RF electric field (see Supplementary Fig. S3).

The amplitude of the peaks in Fig. 2b increases linearly with RF power ($\sim B_{ac}^2$) before saturation occurs, as predicted[25] (Fig. 2b, inset). The ESR satellite peak is expected to be broadened by either the excitation amplitude $B_{ac}$ or incoherent processes, like cotunnelling, inelastic transitions (to the S(0,2) state) or the statistical fluctuations in the nuclear field, whichever of the four has the largest contribution. No dependence of the width on RF power was found within the experimentally accessible range ($B_{ac}$<2 mT). Furthermore, we suspect that the broadening is not dominated by cotunnelling or inelastic transitions because the corresponding rates are smaller than the observed broadening (see Supplementary Figs S4b and S2d). The observed ESR peaks are steeper on the flanks and broader than expected from the nuclear field fluctuations. In many cases, the peak width and position are even hysteretic in the sweep direction, suggesting that the resonance condition is shifted during the field sweep. We speculate that dynamic nuclear polarization due to feedback of the electron transport on the nuclear spins plays a central part here[37].

**Coherent Rabi oscillations**

Following the observation of magnetically induced spin flips, we next test whether we can also coherently rotate the spin by applying RF bursts with variable length. In contrast to the continuous-wave experiment, where detection and spin rotation occur at the same time, we pulse the system into Coulomb blockade during the spin manipulation. This eliminates decoherence induced by tunnel events from the left to the right dot during the spin rotations. The experiment consists of three stages (Fig. 3): initialization through spin blockade in a statistical mixture of ↑↑ and ↓↓, manipulation by a RF burst in Coulomb blockade, and detection by pulsing back for projection (onto S(0,2)) and tunnelling to the lead. When one of the electrons is rotated over (2*n*+1)π (with integer *n*), the two-electron state evolves to ↑↓ (or ↓↑), giving a maximum contribution to the current (as before, when the two spins are anti-parallel, one electron charge moves through the dots).

However, no electron flow is expected after rotations of $2\pi n$, where one would find two parallel spins in the two dots after the RF burst.

We observe that the dot current oscillates periodically with the RF burst length (Fig. 4). This oscillation indicates that we performed driven, coherent electron spin rotations, or Rabi oscillations. A key characteristic of the Rabi process is a linear dependence of the Rabi frequency on the RF burst amplitude, $B_{ac}$ ($f_{Rabi}=g\mu_B B_1/h$ with $B_1=B_{ac}/2$ due to the rotating wave approximation). We verify this by extracting the Rabi frequency from a fit of the current oscillations of Fig. 4b with a sinusoid, which gives the expected linear behaviour (Fig. 4b, inset). From the fit we obtain $B_{ac}$=0.59 mT for a stripline current $I_{CPS}$ of ~1 mA, which agrees well with predictions from numerical finite element simulations (see Supplementary Fig. S1). The maximum $B_1$ we could reach in the experiment before electric field effects hindered the measurement was 1.9 mT, corresponding to $\pi/2$ rotations of only 27 ns (that is, a Rabi period of 108 ns, see Fig. 4b). If the accompanying electric fields from the stripline excitation could be reduced in future experiments (for example, by improving the impedance matching from coax to CPS), considerably faster Rabi flopping should be attainable.

The oscillations in Fig. 4b remain visible throughout the entire measurement range, up to 1 μs. This is striking, because the Rabi period of ~100 ns is much longer than the time-averaged coherence time $T_2^*$ of 10–20 ns (refs 14, 19, 35, 36) caused by the nuclear field fluctuations. The slow damping of the oscillations is only possible because the nuclear field fluctuates very slowly compared to the timescale of spin rotations and because other mechanisms, such as the spin-orbit interaction, disturb the electron spin coherence only on even longer timescales[13,41,42]. We also note that the decay is not exponential (grey line in Fig. 4a), which is related to the fact that the nuclear bath is non-markovian (it has a long memory)[43].

**Theoretical model**

To understand better the amplitudes and decay times of the oscillations, we model the time evolution of the spins throughout the burst duration. The model uses a hamiltonian that includes the Zeeman splitting for the two spins and the RF field, which we take to be of equal amplitude in both dots ($\mathbf{S}_L$ and $\mathbf{S}_R$ refer to the electron spins in the left and right dot respectively):

$H=g\mu_B(\mathbf{B}_{ext}+\mathbf{B}_{L,N})\mathbf{S}_L+g\mu_B(\mathbf{B}_{ext}+\mathbf{B}_{R,N})\mathbf{S}_R+g\mu_B\cos(\omega t)\mathbf{B}_{ac}(\mathbf{S}_L+\mathbf{S}_R)$,

where $\mathbf{B}_{L,N}$ and $\mathbf{B}_{R,N}$ correspond to a single frozen configuration of the nuclear field in the left and right dot. This is justified because the electron spin dynamics is much faster than the dynamics of the nuclear system. From the resulting time evolution operator and assuming that the initial state is a statistical mixture of ↑↑ and ↓↓, we can numerically obtain the probability for having anti-parallel spins after the RF burst. This is also the probability that the left electron tunnels to the right dot during the read-out stage.

In the current measurements of Fig. 4a, each data point is averaged over 15 s, which presumably represents an average over many nuclear configurations. We include this averaging over different nuclear configurations in the model by taking 2,000 samples from a gaussian distribution of nuclear fields (with standard deviation $\sigma=\sqrt{\langle B_N^2\rangle}$), and computing the probability that an electron tunnels out after the RF burst. When the electron tunnels, one or more additional electrons, say *m*, may subsequently tunnel through before ↑↑ or ↓↓ is formed and the current is blocked again. Taking *m* and $\sigma$ as fitting parameters, we find good agreement with the data for *m*=1.5 and $\sigma$=2.2 mT (solid black lines in Fig. 4a. This value for $\sigma$ is comparable to that found in refs 35 and 36. The value found for *m* is different from what we would expect from a simple picture where all four spin states are formed with equal probability during the initialization stage, which would give *m*=1. We do not understand this discrepancy, but it could be due to different tunnel rates for ↑ and ↓ or more subtle details in the transport cycle that we have neglected in the model.

**Time evolution of the spin states during RF bursts**

We now discuss in more detail the time evolution of the two spins during a RF burst. The resonance condition in each dot depends on the effective nuclear field, which needs to be added vectorially to $B_{\text{ext}}$. Through their continuous reorientation, the nuclear spins will bring the respective electron spins in the two dots on and off resonance as time progresses.

When a RF burst is applied to two spins initially in ↑↑, and is on-resonance with the right spin only, the spins evolve as:

$$|\uparrow\rangle|\uparrow\rangle \quad \to \quad |\uparrow\rangle\frac{|\uparrow\rangle+|\downarrow\rangle}{\sqrt{2}} \quad \to \quad |\uparrow\rangle|\downarrow\rangle \quad \to \quad |\uparrow\rangle\frac{|\uparrow\rangle-|\downarrow\rangle}{\sqrt{2}} \quad \to \quad |\uparrow\rangle|\uparrow\rangle$$

When the RF burst is on-resonance with both spins, the time evolution is:

$$|\uparrow\rangle|\uparrow\rangle \quad \to \quad \frac{|\uparrow\rangle+|\downarrow\rangle}{\sqrt{2}}\frac{|\uparrow\rangle+|\downarrow\rangle}{\sqrt{2}} \quad \to \quad |\downarrow\rangle|\downarrow\rangle \quad \to \quad \frac{|\uparrow\rangle-|\downarrow\rangle}{\sqrt{2}}\frac{|\uparrow\rangle-|\downarrow\rangle}{\sqrt{2}} \quad \to \quad |\uparrow\rangle|\uparrow\rangle$$

In both cases, the RF causes transitions between the ↑ and ↓ states of single spin-half particles. When the RF is on-resonance with both spins, such single-spin rotations take place for both spins simultaneously. Because the current through the dots is proportional to the $S_z$=0 probability (↑↓ or ↓↑), we see that when both spins are excited simultaneously, the current through the dots will oscillate twice as fast as when only one spin is excited, but with only half the amplitude.

In the experiment, the excitation is on-resonance with only one spin at a time for most of the frozen nuclear configurations (Fig. 5). Only at the highest powers ($B_1/\sqrt{\langle B_{\text{N},z}^2 \rangle} > 1$), both spins may be excited simultaneously (but independently) and a small double Rabi frequency contribution is expected, although it could not be observed, owing to the measurement noise.

**Quantum gate fidelity**

We can estimate the angle over which the electron spins are rotated in the Bloch sphere based on our knowledge of $B_1$ and the nuclear field fluctuations in the *z*-direction, again using the hamiltonian *H*. For the maximum ratio of $B_1/\sqrt{\langle B_{\text{N},z}^2 \rangle} = B_1/(\sigma/\sqrt{3}) = 1.5$ reached in the present experiment, we achieve an average tip angle of 131° for an intended 180° rotation, corresponding to a fidelity of 73% (Fig. 5). Apart from using a stronger $B_1$, the tip angle can be increased considerably by taking advantage of the long timescale of the nuclear field fluctuations. First, application of composite pulses, widely used in nuclear magnetic resonance to compensate for resonance off-sets[44], can greatly improve the quality of the rotations. A second solution comprises a measurement of the nuclear field (nuclear state narrowing[45–47]), so that the uncertainty in the nuclear field is reduced, and accurate rotations can be realized for as long as the nuclear field remains constant.

In future experiments, controllable addressing of the spins in the two dots separately can be achieved through a gradient in either the static or the oscillating magnetic field. Such gradient fields can be created relatively easily using a ferromagnet or an asymmetric stripline. Alternatively, the resonance frequency of the spins can be selectively shifted using local *g*-factor engineering[48,49]. The single spin rotations reported here, in combination with single-shot spin read-out[13,18] and the tunable exchange coupling in double dots[14], offers many new opportunities, such as measuring the violation of Bell's inequalities or the implementation of simple quantum algorithms.

**Acknowledgements**

We thank W. Coish, J. Elzerman, D. Klauser, A. Lupascu, D. Loss and in particular J. Folk for discussions; R. Schouten, B. van der Enden and W. den Braver for technical assistance; The International Research Centre for Telecommunication and Radar at the Delft University of Technology for assistance with the stripline simulations. Supported by the Dutch Organization for Fundamental Research on Matter (FOM), the Netherlands Organization for Scientific Research (NWO) and the Defense Advanced Research Projects Agency Quantum Information Science and Technology programme.

**Author information**

Correspondence and requests for materials should be addressed to L.M.K. Vandersypen (l.m.k.vandersypen@tudelft.nl) and F.H.L. Koppens (f.h.l.koppens@tudelft.nl).

# Figure 1

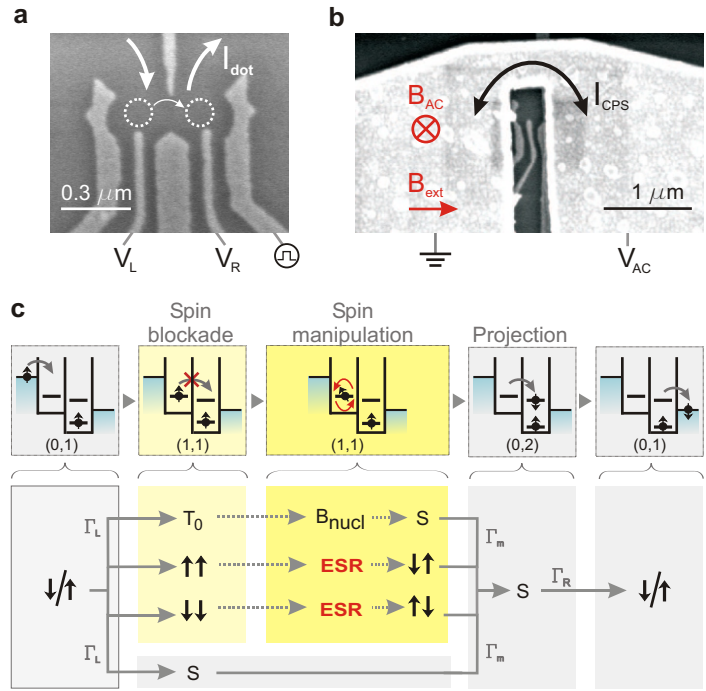

**Device and ESR detection scheme.**

**a**, Scanning electron microscope (SEM) image of a device with the same gate pattern as used in the experiment. The Ti/Au gates are deposited on top of a GaAs/AlGaAs heterostructure containing a two-dimensional electron gas 90 nm below the surface. White arrows indicate current flow through the two coupled dots (dotted circles). The right side gate is fitted with a homemade bias-tee (rise time 150 ps) to allow fast pulsing of the dot levels.

**b**, SEM image of a device similar to the one used in the experiment. The termination of the coplanar stripline is visible on top of the gates. The gold stripline has a thickness of 400 nm and is designed to have a 50 $\Omega$ characteristic impedance, $Z_0$, up to the shorted termination. It is separated from the gate electrodes by a 100-nm-thick dielectric (Calixerene)[50].

**c**, Diagrams illustrating the transport cycle in the spin blockade regime. This cycle can be described via the occupations (*m,n*) of the left and right dots as (0,1) → (1,1) → (0,2) → (0,1). When an electron enters the left dot (with rate $\Gamma_L$) starting from (0,1), the two-electron system that is formed can be either a singlet S(1,1) or a triplet T(1,1). From S(1,1), further current flow is possible via a transition to S(0,2) (with rate $\Gamma_m$). When the system is in T(1,1), current is blocked unless this state is coupled to S(1,1). For $T_0$, this coupling is provided by the inhomogeneous nuclear field $\Delta B_N$. For $T_+$ or $T_-$, ESR causes a transition to ↓↑ or ↑↓, which contains a S(1,1) component and a $T_0$ component (which is in turn coupled to S(1,1) by the nuclear field).

# Figure 2

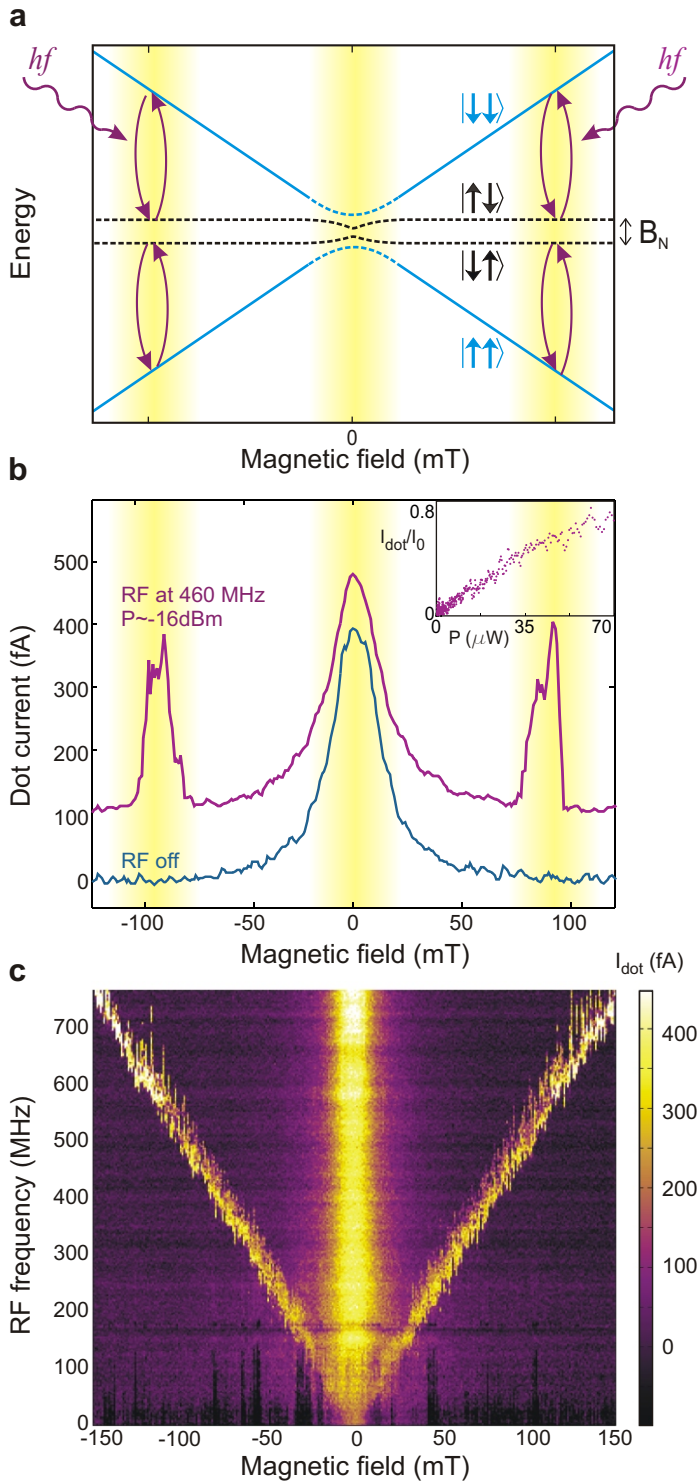

**ESR spin state spectroscopy.**

**a**, Energy diagram showing the relevant eigenstates of two electron spins in a double-dot, subject to an external magnetic field and nuclear fields. Because the nuclear field is generally inhomogeneous, the Zeeman energy is different in the two dots and results therefore in a different energy for and . ESR turns the spin states and into or , depending on the nuclear fields in the two dots. The yellow bands denote the ranges in $B_{ext}$ where spin blockade is lifted (by the nuclear field or ESR) and current will flow through the dots.

**b**, Current measured through the double-dot in the spin blockade regime, with (red trace, offset by 100 fA for clarity and without (blue trace) a RF magnetic field. Satellite peaks appear as the external magnetic field is swept through the spin resonance condition. Each measurement point is averaged for one second, and is therefore expected to represent an average response over many nuclear configurations. The RF power $P$ applied to the CPS is estimated from the power applied to the coax line and the attenuation in the lines. Inset, satellite peak height versus RF power ($f$=408 MHz, $B_{ext}$=70 mT, taken at slightly different gate voltage settings). The current is normalized to the current at $B_{ext}$=0 ($=I_0$). Unwanted electric field effects are reduced by applying a compensating signal to the right side gate with opposite phase as the signal on the stripline (see Supplementary Fig. S4). This allowed us to obtain this curve up to relatively high RF powers.

**c**, Current through the dots when sweeping the RF frequency and stepping the magnetic field. The ESR satellite peak is already visible at a small magnetic field of 20 mT and RF excitation of 100 MHz, and its location evolves linearly in field when increasing the frequency. For higher frequencies the satellite peak is broadened asymmetrically for certain sweeps, visible as vertical stripes. This broadening is time dependent, hysteretic in sweep direction, and changes with the dot level alignment. The horizontal line at 180 MHz is due to a resonance in the transmission line inside the dilution refrigerator.

# Figure 3

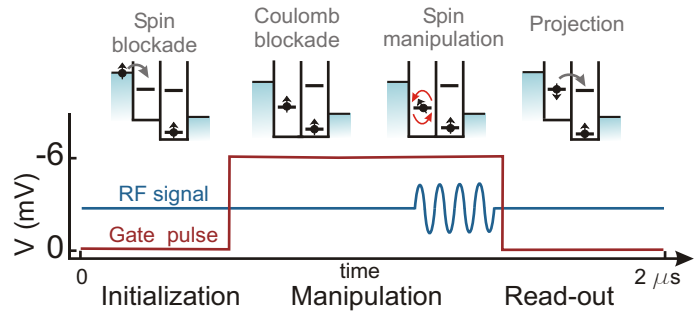

**The control cycle for coherent manipulation of the electron spin.**

During the 'initialization' stage the double-dot is tuned in the spin blockade regime. Electrons will move from left to right until the system is blocked with two parallel spins (either ↑↑ or ↓↓; in the figure only the ↑↑ case is shown). For the 'manipulation' stage, the right dot potential is pulsed up so none of the levels in the right dot are accessible (Coulomb blockade), and a RF burst with a variable duration is applied. 'Read-out' of the spin state at the end of the manipulation stage is done by pulsing the right dot potential back; electron tunnelling to the right lead will then take place only if the spins were anti-parallel. The duration of the read-out and initialization stages combined was 1 μs, long enough (1 μs ≫ 1/$\Gamma_L$, 1/$\Gamma_M$, 1/$\Gamma_R$) to have parallel spins in the dots at the end of the initialization stage with near certainty (this is checked by signal saturation when the pulse duration is prolonged). The duration of the manipulation stage is also held fixed at 1 μs to keep the number of pulses per second constant. The RF burst is applied just before the read-out stage starts.

# Figure 4

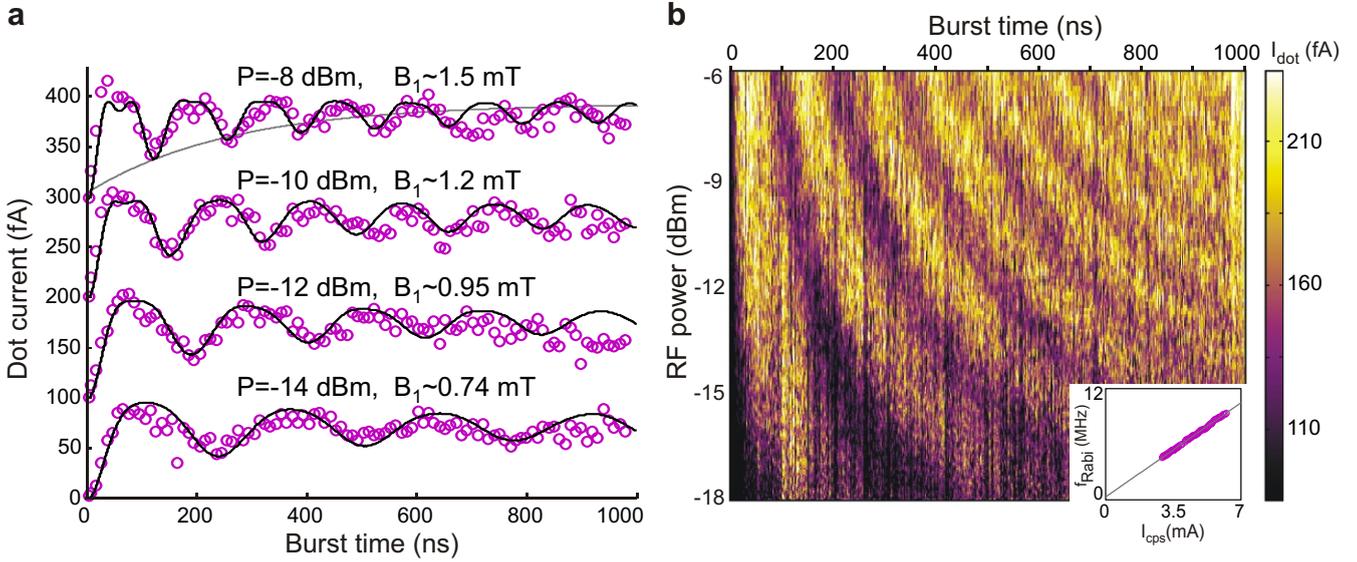

**Coherent spin rotations.**

**a**, The dot current—reflecting the spin state at the end of the RF burst—oscillates as a function of RF burst length (curves offset by 100 fA for clarity). The frequency of $B_{ac}$ is set at the spin resonance frequency of 200 MHz ($B_{ext}$=41 mT). The period of the oscillation increases and is more strongly damped for decreasing RF power. The RF power $P$ applied to the CPS is estimated from the power applied to the coax line and the attenuation in the lines and RF switch. From $P$, the stripline current is calculated via the relation $P = \frac{1}{2}\left(\frac{I_{CPS}}{2}\right)^2 Z_0$ assuming perfect reflection of the RF wave at the short. Each measurement point is averaged over 15 s. We correct for a current offset which is measured with the RF frequency off-resonance (280 MHz). The solid lines are obtained from numerical computation of the time evolution, as discussed in the text. The grey line corresponds to an exponentially damped envelope.

**b**, The oscillating dot current (represented in colourscale) is displayed over a wide range of RF powers (the sweep axis) and burst durations. The dependence of the Rabi frequency $f_{Rabi}$ on RF power is shown in the inset. $f_{Rabi}$ is extracted from a sinusoidal fit with the current oscillations from 10 to 500 ns for RF powers ranging from -12.5 dBm up to -6 dBm.

# Figure 5

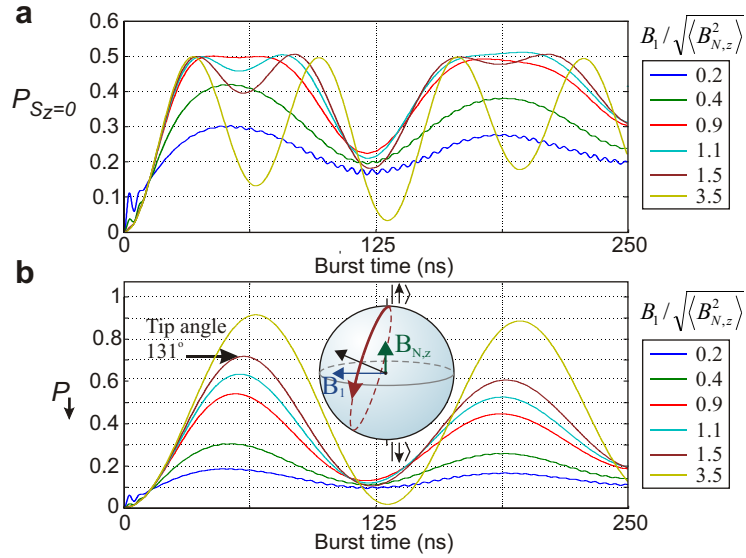

**Time evolution of the spin states.**

**a**, Probability for the two spins to be in ↑↓ or ↓↑ ($S_z=0$) at the end of a RF burst, with initial state ↑↑, computed using the hamiltonian $H$ presented in the main text, for six different values of $B_{N,z} \equiv \langle B_{N,z}^2 \rangle^{1/2}$ (fixed $B_1=1.5$ mT, $B_{ext}=40$ mT, each of the traces is averaged over 2,000 static nuclear configurations). As expected, the oscillation contains a single frequency for $B_1$ small compared to $B_{N,z}$, corresponding to the Rabi oscillation of a single spin. The oscillation develops a second frequency component, twice as fast as the first, when $B_1/B_{N,z}>1$. For $B_1/B_{N,z}>4$ the double frequency component is dominant, reflecting the simultaneous Rabi oscillation of the two spins.

**b**, Probability for one of the spins to be ↓ at the end of a RF burst. The spin state evolution is computed as in **a**. This oscillation represents the Rabi oscillation of one spin by itself. For increasing $B_1$, the maximum angle over which the spin is rotated in the Bloch sphere increases as well. In the experiment, this angle could not be measured directly, because the current measurement constitutes a two-spin measurement, not a single-spin measurement. We can, however, extract the tip angle from $P_↓$.

# Driven coherent oscillations of a single electron spin in a quantum dot

F.H.L. Koppens, C. Buizert, K.J. Tielrooij, I.T. Vink, K.C. Nowack, T. Meunier, L.P. Kouwenhoven, L.M.K. Vandersypen

## Supplementary on-line material

**Supplementary Fig. S1: Coplanar stripline design**

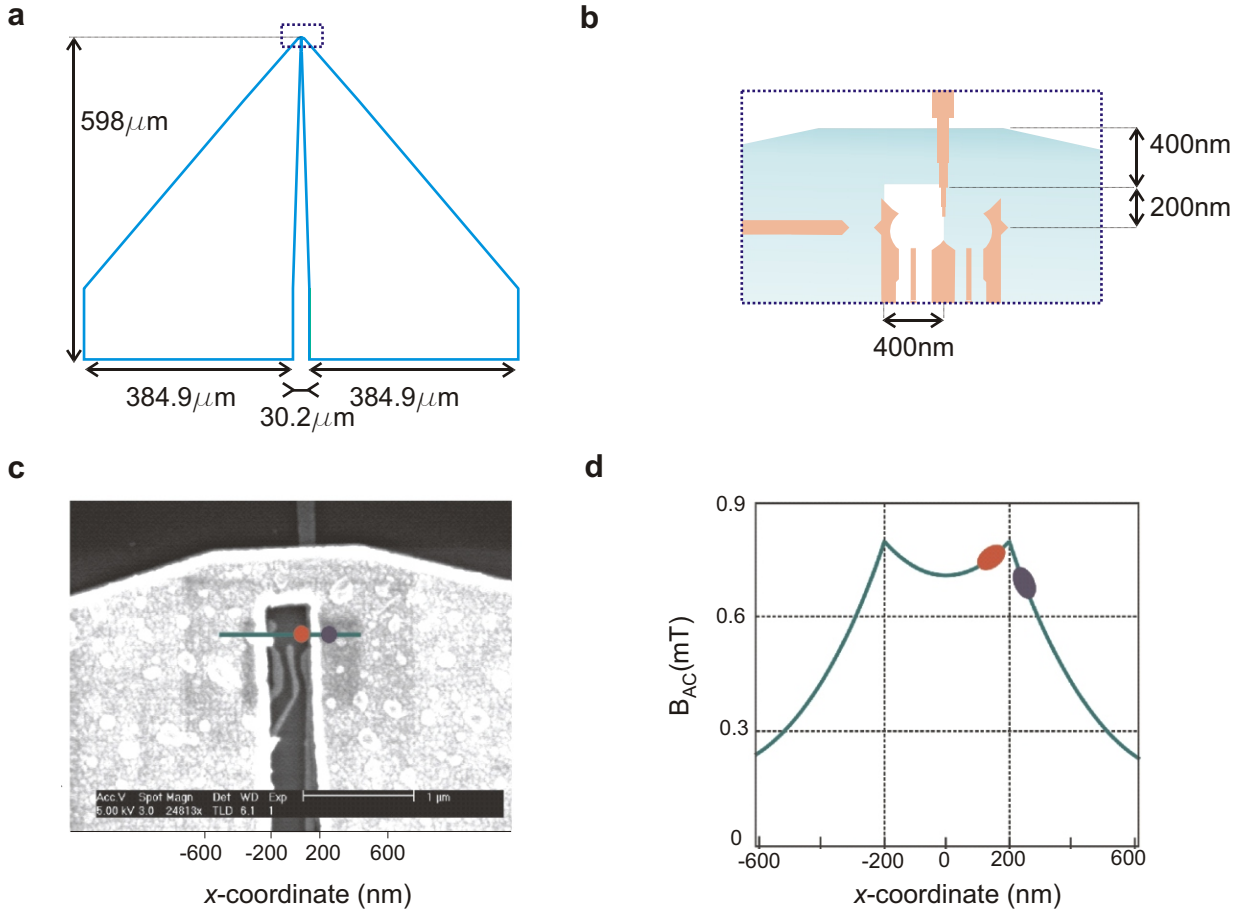

**a** Schematic diagram of the on-chip coplanar stripline. The CPS is terminated by a narrow wire that shorts the two planes. The wire effectively acts as a shorted termination of the 50 Ω transmission line and therefore the current will exhibit an anti-node at the wire.

**b** Schematic diagram showing the termination of the stripline and the position of the surface gates that define the double quantum dot. The design is optimized to maximize $B_{ac}$ at the location of the dots.

**c** SEM image of a device similar to the one used in the experiment. The termination of the CPS is visible as well as part of the surface gates that define the dots. The estimated locations of the two quantum dots are indicated in red and blue.

**d** Amplitude of the oscillating magnetic field perpendicular to the plane, 200 nm below the CPS, along the green line in **c** (P=-22 dBm, f=*200* MHz), computed numerically using CST Microwave Studio. This program solves the integral form of Maxwell's equations with the finite difference time domain method for a discretised spatial domain. In the simulation, an ideal waveguide source is connected to the CPS, through which a quasi-TEM wave will propagate. The approximate x-coordinates of the dots are indicated in blue and red. Based on these simulation results, we expect a field of $B_{ac}$~0.7 mT for a -22 dBm excitation (corresponding to $I_{cps}$~1 mA) at 200 MHz. Furthermore, we expect the fields in the two dots to differ from each other by no more than 20%.

**Supplementary Fig. S2: Spin blockade and leakage current**

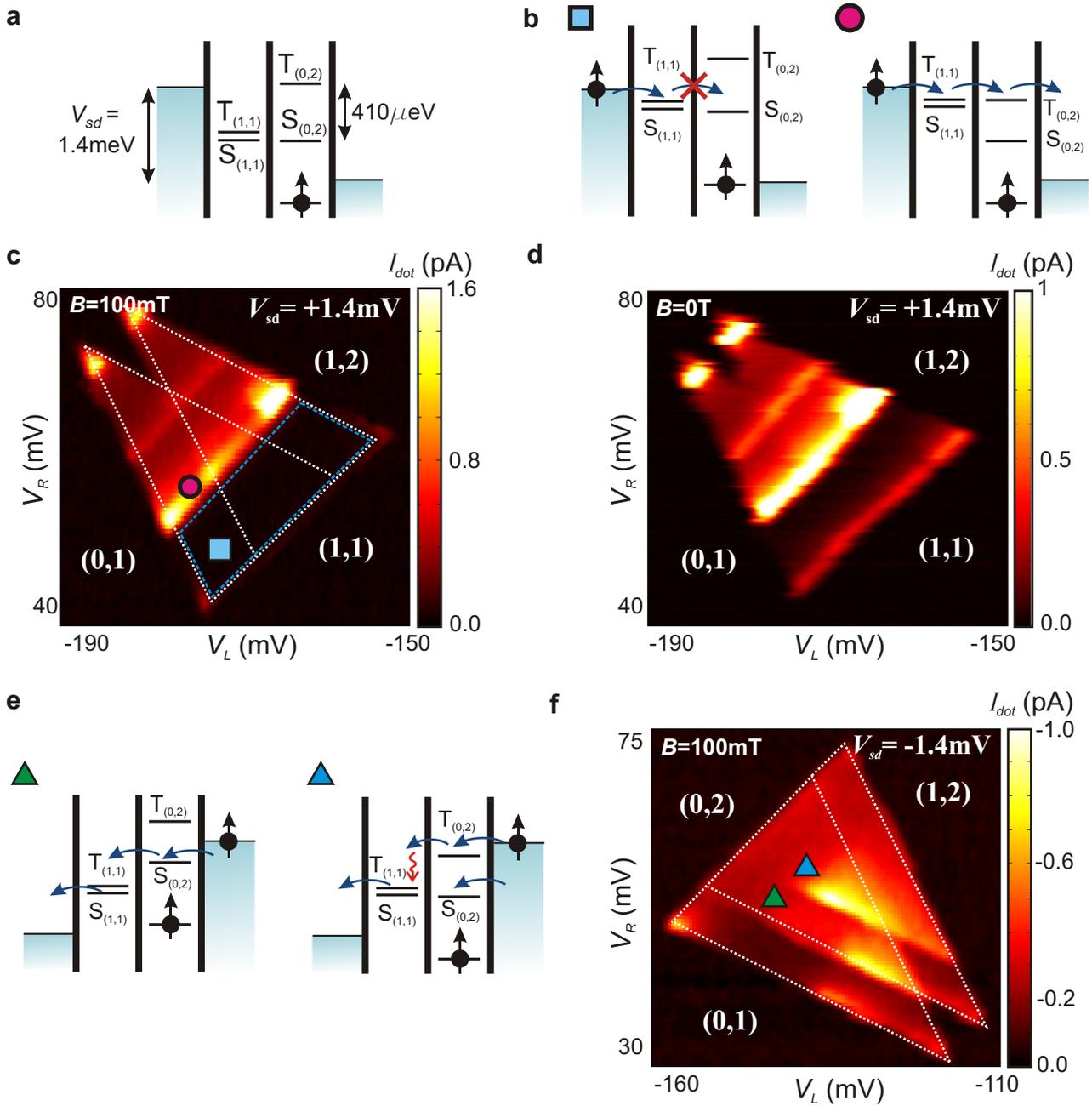

**a** Schematic of a double quantum dot tuned to the Pauli spin blockade regime[30,34]. The spin blockade makes use of the fact that the single dot singlet-triplet splitting, between S(0,2) and T(0,2), is much larger than the double dot singlet-triplet splitting, between S(1,1) and T(1,1).

**b** (Left) When an electron tunnels to the left dot from the reservoir and the T(1,1) state is formed, the electron cannot transition to the right dot because the T(0,2) state is not energetically accessible. (Right) When the right dot potential is pulled down, the T(0,2) becomes accessible and spin blockade is lifted.

**c** The current measured through the double dot under forward bias is plotted in colorscale, as a function of the gate voltages controlling the left and right dot potential ($B_{ext}$ = 100 mT). The white dotted triangles define the region in gate space where transport is energetically allowed51 (outside these triangles, the number of electrons is fixed by Coulomb blockade). In part of the triangles, transport is still suppressed, due to spin blockade (blue dotted lines). The color markers refer to the diagrams in b. The white numbers indicate the number of electrons in the left and right dot, based on transport measurement through the double-dot. The dot occupation could not be verified explicitly with QPC measurements leaving ambiguity about the absolute electron number.

**d** Similar measurement as in **c** but for $B_{ext}$ = 0 T. At zero external field, the nuclear field admixes all the triplets with the singlet, and spin blockade is lifted. This is only possible when the energy splitting J between S(1,1) and T(1,1) is smaller than $\sqrt{\langle B_N^2 \rangle}$ [36], which is the case here. This was also verified via the field dependence of the leakage current at $\Delta_{LR}$=0 (resonance transport) around $B_{ext}$ = 0 T. We found no splitting of the leakage current peak on the field axis which indicates $\sqrt{2}t \gtrsim J \lesssim g\mu_B \sqrt{\langle B_N^2 \rangle}$ [36]. At finite field, the blockade can be lifted by ESR (not shown). The upper bound for $t$ allows us to extract a bound for inelastic transitions $\Gamma_{in}$ to the S(0,2) state: $h\Gamma_{in} \lesssim g\mu_B \sqrt{\langle B_N^2 \rangle}$.

**e** Schematic of a double dot in similar conditions as in **a** and **b** but with the bias voltage across the dot reversed. The system can never be stuck in T(1,1) now, so spin blockade does not occur.

**f** Similar measurement as in **c**, but for reverse bias. As expected, current flows in the entire region in gate space where it is energetically allowed (within the white dotted triangles).

**Supplementary Fig. S3: Verification of magnetic excitation**

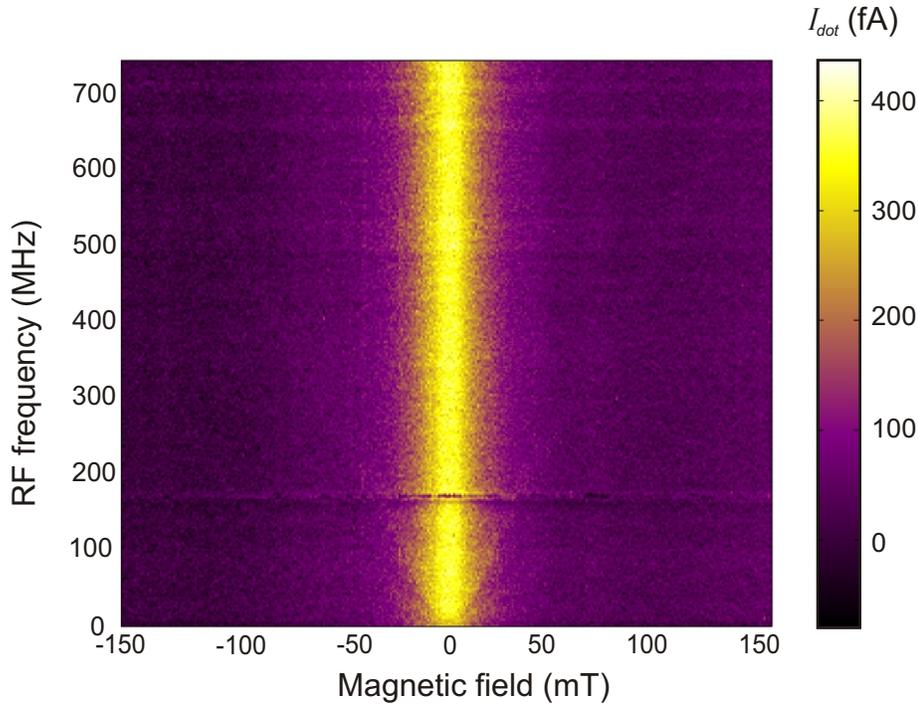

Similar data as in Fig. 2c of the main article, but now with the RF signal applied to the right side gate instead of to the ESR stripline. Ideally, the excitation should now be purely electric. The amplitude of the RF signal (-50 dBm at the gate) was chosen such that the electric field is equally strong as in the ESR measurements of Fig. 2c (this was determined from the measured PAT rate). The satellite peaks that were clearly visible in Fig. 2c have (almost) disappeared. Nevertheless, a very faint line is still present at the same position as the ESR response in Fig. 2c. This response could be due to the small magnetic field generated by the current in the gate, which is capacitively coupled to its environment. It could also be due to the coupling of the electric field to the electron spin, through Rashba or Dresselhaus spin-orbit interaction[28]. A final possibility is that spins are flipped when the electron wave function is moved back-and-forth in the inhomogeneous nuclear field[29]. In any case, it is clear that in our experiment, all these mechanisms are much less efficient than magnetic excitation via the CPS.

**Supplementary Fig. S4: Photon-assisted tunneling due to electric fields**

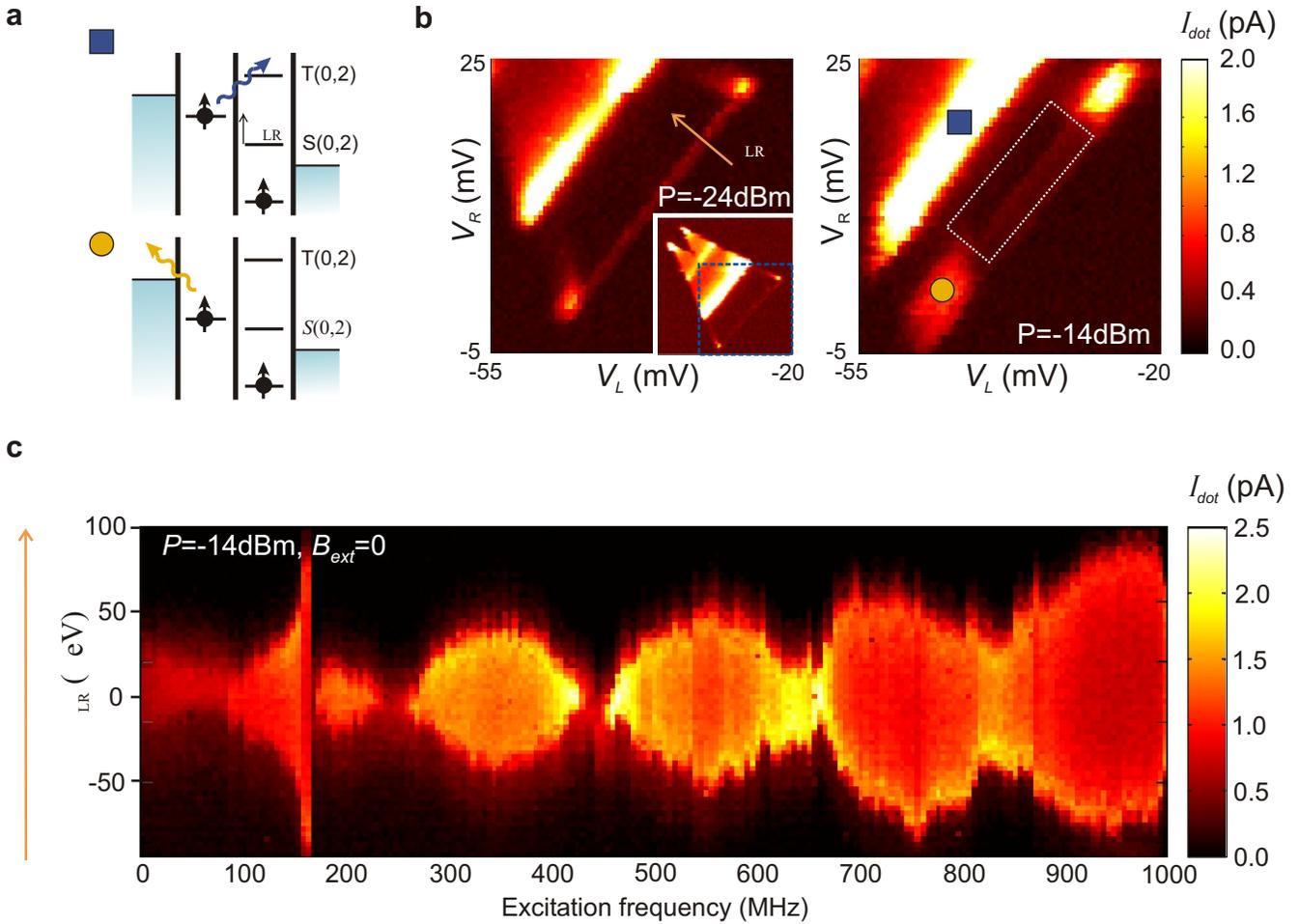

**a** Diagrams showing schematically two photon-assisted tunneling (PAT) processes. Top: excitation from T(1,1) to T(0,2). Bottom: excitation from T(1,1) to the left reservoir.

**b** Similar measurements as in Fig. S2c with RF power -24 dBm and -14 dBm applied to the CPS ($B_{ext}$=100 mT). The effect of the two PAT processes on the measured current is visible as current enhancement in the areas around the yellow circle and blue square. ESR detection in the experiments discussed in the main text has been performed in the area enclosed by the white dotted lines where the PAT rates are smaller than the measurement noise (~10-100fA).

**c** Current as a function of CPS excitation frequency and $\Delta_{LR}$ (energy difference between S(1,1) and S(0,2) state, see also **a** and **b**) with an RF signal applied both to the right side gate (time-delayed and 34 dB attenuated) and the CPS. The amplitude of the total electric field, reflected in the broadening of the current peak along the vertical axis, shows constructive and destructive interference as a function of frequency.

# B. Supplementary Discussion

**Photon assisted tunneling due to electric fields**

The coplanar stripline is designed to maximize the ratio between the RF magnetic field and electric field. Nevertheless, a small RF electric field will unavoidably be generated. High frequency electric fields can excite an electron to higher lying orbitals in the dot or in the reservoir. In this process, one or more photons are absorbed to match the excitation energy. Such so-called photon-assisted tunneling (PAT) processes[52,53] can lift spin blockade and overwhelm the ESR signal.

In this section we will discuss two different kinds of PAT processes that can lift spin blockade (Figs. S4a,b). The first is PAT through the interdot barrier. Electrons blocked in any of the three T(1,1) states can tunnel to the T(0,2) state if the energy difference between these states corresponds to an integer multiple of the photon energy *hf*. This will lead to sideband resonances running parallel to the T(0,2) line with a spacing *hf*. In the classical limit, where *hf* is much smaller than the line width of the states *hΓ* (*Γ* is the tunnel rate), the individual sidebands cannot be resolved. Instead the T(0,2) line is broadened.

We can quantify how efficient PAT is in lifting spin blockade using ref.[52]. The basic idea is that an AC voltage drop $V=V_{ac}cos(2\pi ft)$ across a tunnel barrier modifies the tunnel rate through the barrier as $\tilde{\Gamma}(E) = \sum_{n=-\infty}^{+\infty} J_n^2(\alpha)\Gamma(E+nhf)$. Here $\Gamma(E)$ and $\tilde{\Gamma}(E)$ are the tunnel rates at energy *E* with and without an AC voltage, respectively; $J_n^2(\alpha)$ is the square of the *n*th order Bessel function evaluated at $\alpha = (eV_{ac})/hf$, which describes the probability that an electron absorbs or emits *n* photons of energy *hf* (*-e* is the electron charge). The energy splitting between the S(0,2) and T(0,2) states is typically ~410 µeV, and the energy difference between the T(1,1) and the T(0,2) state will be of the same order. Since we can keep the Zeeman splitting small in this double dot measurement, the excitation frequency can be kept small too. Typically, *f* =200 MHz in the present experiment. The single-photon energy is then *hf*=0.8 µeV. PAT processes from T(1,1) to T(0,2) thus require *n*=500 photons, and will therefore be very inefficient. Such 500-photon processes only occur with a reasonable probability, $J_n^2(\alpha)>0.05$, if α > *n*-1~500. So only when the amplitude of the oscillating voltage across the central barrier exceeds roughly 400 µV, spin blockade is lifted due to PAT from the T(1,1) to the T(0,2) state. In the continuous-wave experiment this occurs for RF powers larger than -12 dBm.

The second PAT process occurs through the outer barriers. The electron blocked in the left dot can be excited to the left reservoir if the Fermi level of this reservoir lies within *nhf* in energy from the T(1,1) electrochemical potential (Fig. S4a). Subsequently, another electron with possibly a different spin state can tunnel from the left reservoir into the dot. Effectively this process can thus flip the spin by electron exchange with the reservoir. Similarly, the electron in the right dot can be excited to the right reservoir. The data presented in Figs. 2 and 4 of the main Article are taken with a large bias voltage of 1.4 mV applied across the double dot, and with the relevant levels in the left and right dots far separated in energy from the Fermi level in the corresponding reservoir. In this way, PAT processes to the reservoirs were minimized.

We point out that a third process, namely photon-assisted tunneling from the S(1,1) state to the S(0,2) state, does not disturb ESR detection. This process only broadens the ESR peak on the gate axis (defined by $\Delta_{LR}$, see Figs. S4a,b).

Even though PAT can thus be easily recognized and minimized in double dot measurements, PAT rates still became excessive at higher RF powers. This imposed a limitation on the power we could apply to the CPS, and thus on the amplitude of $B_{ac}$ we could produce in the experiment (before heating of the sample or the mixing chamber became a limitation). We therefore developed a method to reduce the PAT rates via interference between two signals. Hereby we split the RF signal at the output of the source, send one branch directly to the CPS and send the other branch to the right side gate of the dot. The latter signal is delayed through an additional coax of length $L$ and properly attenuated such that for specific RF frequencies $f=(n+½)c/\Delta L$, the electric field generated by the CPS interferes destructively with the electric field created by the side gate (Fig. S4c). At the frequencies that correspond to nodes in the interference pattern, it is possible to apply about 6 dB more RF power than is possible without PAT cancellation.

Only the data in the inset of Fig. 2b of the main Article are obtained with PAT reduction. For the pulsed experiments, the PAT rate to the T(0,2) state is smaller than in the continuous-wave experiments, because the right dot levels are pulsed to a higher energy (thereby increasing the energy difference between T(0,2) and T(1,1)) when the microwaves are applied.

## C. Supplementary methods

The RF signals were generated using a HP 83650A source, which covers a frequency range from 10 MHz to 50 GHz, with 1 kHz resolution. RF bursts were created by sending this signal through a high isolation GaAs RF switch (Minicircuits ZASWA-2-50DR, typical rise time 3 ns), gated by rectangular pulses from an arbitrary waveform generator (Tektronix AWG520, marker channel, rise time ≤ 2.5 ns). The coplanar stripline is contacted via a modified microwave probe (GGB Industries, Picoprobe model 50A, loss <1 dB, DC-50 GHz).

Voltage pulses are applied to the right side gate through a bias-tee, so that the gate can remain biased with a DC voltage as well. The bias-tee was home-made, with a rise time of 150 ps and a RC charging time of »10ms at 77K (R=10 MΩ, C= 3.3nF). The voltage pulses were generated using the arbitrary waveform channel of the same source as used to gate the RF switch.

The GaAs/AlGaAs heterostructure from which the samples were made was purchased from Sumitomo Electric. The 2DEG has a mobility of $185 \times 10^3$ cm$^2$/Vs at 77 K, and an electron density of 4-5 $\times 10^{11}$ cm$^{-2}$, measured at 30 mK with a different device than used in the experiment. Background charge fluctuations made the quantum dot behaviour excessively irregular. The charge stability of the dot was improved considerably in two ways. First, the gates were biased by +0.5 V relative to the 2DEG during the device cool-down. Next, after the device had reached base temperature, the reference of the voltage sources and I/V converter (connected to the gates and the 2DEG) were biased by +2 V. This is equivalent to a -2 V bias of both branches of the CPS, which therefore (like a gate) reduces the 2DEG density under the CPS.

The measurements were performed in a Oxford Instruments Kelvinox 400 HA dilution refrigerator operating at a base temperature of ~38mK.

## D. Supplementary Notes